\documentclass[onecolumn,showpacs,preprintnumbers,amsmath,amssymb]{revtex4}%

\usepackage{epsfig}%
\usepackage{graphicx}
\usepackage{dcolumn}
\usepackage{bm}
\begin{document}
\title{Brain, Music and non-Poisson Renewal Processes}
\author{Simone Bianco$^1$ }
\author{Massimiliano Ignaccolo$^{1}$}
\author{Mark S. Rider$^{2}$}
\author{Mary J. Ross$^3$}
\author{Phil Winsor$^{4}$}
\author{Paolo Grigolini$^{1,5,6}$}
\affiliation{$^1$Center for Nonlinear Science, University of North Texas,
    P.O. Box 311427,Denton, Texas 76203-1427, USA}
\affiliation{$^2$Integrative and Health Psychology, $700$ W. Wall st., suite
    $200$, Grapevine, Tx, 76051}
\affiliation{$^3$Neurotherapy Lab, Department of Rehabilitation, Social Work
  and Addiction, University of North Texas, P.O. Box 310829, Denton, Texas 76203-0829}
\affiliation{$^4$Center for Audio-Visual Experiment Microelectronics and
    Information Systems Research Center National Chiao-Tung University 1001 Ta
    Hsueh Road Hsin-Chu, Taiwan 30050}
\affiliation{$^5$Istituto dei Processi Chimico Fisici del CNR, Area della
    Ricerca di Pisa, Via G. Moruzzi, 56124, Pisa, Italy}
\affiliation{$^6$Dipartimento di Fisica "E.Fermi" - Universit\`{a} di
Pisa, Largo  Pontecorvo, 3 56127 PISA}

\date{\today}
\begin{abstract}
In this paper we show that both music composition and brain function, as
revealed by the Electroencephalogram (EEG) analysis,  are renewal non-Poisson
processes living in the non-ergodic dominion. To reach this important
conclusion we process the data with the minimum spanning tree method, so as to
detect significant events, thereby building a sequence of times, which is the
time series to analyze. Then we show that in both cases, EEG and music
composition, these significant events are the signature of a non-Poisson
renewal process. This conclusion is reached using a technique
of statistical analysis recently developed by our group, the Aging Experiment
(AE). First, we find that in both cases
the distances between two consecutive events are described by non-exponential
histograms, thereby proving the non-Poisson nature of these processes. The
corresponding survival probabilities $\Psi(t)$ are well fitted by stretched
exponentials ($\Psi(t) \propto exp(-(\gamma t)^\alpha$), with $0.5
<\alpha <1$.) The
second step rests on the adoption of AE, which shows that these
are renewal processes.
We show that the stretched exponential, due to its renewal character, is the emerging tip of an
iceberg, whose underwater part has slow tails with an inverse power law
structure with power index $\mu = 1 + \alpha$.
Adopting the AE procedure we find that both EEG and
music composition  yield $\mu < 2$.
On the basis of the recently discovered  complexity matching effect, according
to which a complex system $S$ with $\mu_S < 2$ responds only to a complex
driving signal $P$ with $\mu_P \leq \mu_S$, we conclude that the results of our
analysis may  explain the influence of music on the human brain.

      \end{abstract}
\maketitle

\section{Introduction}

The study of neuronal systems is a challenge for statistical physics,
insofar as experimental evidence is proving that the ordinary Poisson
paradigm is inadequate to deal with these complex systems. According
to some neuro-physiologists the neurons are renewal \cite{vreeswijk}
and they are markedly non-Poisson \cite{baddeley}. More precisely, the
experimental evidence of \emph{in vitro} observations coupled with
analysis of \emph{in vivo} spiking patterns indicate that single
neurons are fundamentally non-Poisson processes \cite{notpoisson}. In
the literature the implicit assumption is frequently made that even if
the spiking activity of a single neuron is not Poissonian, the
activity of a set of many neurons is Poissonian. This assumption may
lead us to conclude that the human brain is a Poisson system. The authors
of the paper of Ref.~\cite{remarkablediscovery} proved that this assumption
is invalid. In this paper, through the Electroencephalogram (EEG)
analysis we reach the conclusion that the human brain is not a Poisson
system, in line with the theoretical remarks of
Ref.~\cite{remarkablediscovery}. It is important to point out that in 
this paper the term non-Poisson process indicates a strong deviation 
from the exponential decay. This kind of non-Poisson behavior 
implies, as we shall see, the emergence of fat tails with an inverse 
power law behavior, although for many reasons, ranging from the 
finite time observation to the influence of spurious random 
fluctuations, in the long-time region these tails are truncated.

In this paper we apply the same statistical analysis to music
composition. There is a wide agreement that music composition can be
thought of as a complex signal. It is convenient to quote the seminal
work by Voss and Clarke \cite{vossandclarke}. These authors have found
in fact that music composition yields a $1/f$ noise spectrum, which is
generally regarded as a complexity 
manifestation~\cite{bak,wife,widow,raincomplexity}.

The power of music to evoke emotions is well known, but only recently,
as discussed in the short review of Ref. \cite{andrade}, it has
attracted the attention of neuro-scientists \cite{zatorre}. How does
the communication between music and the brain take place? The authors of
an interesting paper \cite{premusic}
have recently studied the brain response of musicians and
non-musicians to music listening and have found that musicians
yield a higher degree of the gamma band synchrony. A more recent
study of these investigators \cite{morerecent} has supported the
theory that phase synchronization is a significant marker in human
cognition \cite{varela}.

The phase synchronization established by the authors of
Refs.~\cite{premusic,morerecent} is one of the interesting properties
of chaos synchronization~\cite{boccaletti}, which
is in fact attracting the increasing interest of
neuro-scientists~\cite{neuroscientists}.  According to the seminal work of
Ref.~\cite{attractor} both driving system (the music composition) and
response system (the brain) are dynamic systems with strange
attractors (for instance R\"{o}ssler systems \cite{rossler}).

In this paper we analyze both EEG data and music composition with the
joint use of two techniques, the Minimum Spanning Tree (MST) approach and the
Aging Experiment (AE).  Both EEG and music composition data are
expressed as multidimensional vectors $\bf V_{i}$, with $i =
1,\ldots, N$. The different components of these vectors correspond to the
information afforded by different electrodes in the EEG case, and
timber, pitch, harmony, melody, rhythm \ldots etc., in the case of music
composition. In this representation the existence of events is not 
evident. Thus to reveal their presence we build the MST proposed by 
Kruskal \cite{kruskal} and we study its time evolution: From time to 
time the
topology of this MST
undergoes an abrupt change that we interpret as an event. The
probability of an event occurrence at a time distance $\tau$ from the
preceding event is found to be a  stretched exponential. With the help of the AE
method we prove this process to be renewal.
The conclusion of this paper is that both EEG and music composition
are non-Poisson renewal processes, generated by an inverse power law
distribution with index $\mu < 2$.
This condition apparently departs from those conditions considered to be
essential for synchronization, and so for the brain to be sensitive to
music. We note however that it has been shown \cite{letushope,cme}
that while a non-Poisson complex system does not respond to harmonic
perturbation~\cite{doesnotrespond,barbi}, it is very sensitive to the
influence of a complex perturbation with the same complexity. The
authors of Ref.~\cite{letushope,cme} denoted this effect as Complexity Matching
(CM) phenomenon. Therefore, the non-Poisson renewal condition is
compatible with the transmission of information from a complex system
to another, and, on top of that, thanks to the CM effect, might allow us
to explain the influence of music on the brain.

The outline of this paper is as follows. In Section \ref{stretched}, 
with the help of Appendix A,  we show
under which condition a survival probability (SP)  with the form of a stretched
exponential can be considered a non-Poisson renewal process with power law
index $\mu < 2$.  In Section \ref{MST} we illustrate the MST method. Sections
\ref{EEG} and \ref{music} illustrate how we reached the conclusion that both
EEG and music composition data are manifestations of non-Poisson renewal
processes with $\mu < 2$. In Section \ref{CM} we concisely review the CM
effect so as to support the main conclusion, illustrated in Section
\ref{theend}, that it is possible to tune music composition to the brain so as to
make the brain respond to music.

\section{The stretched exponential as a truncated Mittag-Leffler
function} \label{stretched}

The purpose of this Section is to derive 
the following conclusions: 
(i) A particular class of stochastic 
processes leads, in an appropriate limit, to a renewal process with SP 
described by a Mittag-Leffler (ML) function; (ii) The ML function has 
the small $\tau$  limiting form of a stretched exponential, 
$\exp(-(\gamma \tau^{\alpha}))$, with $\alpha < 1$,  and a large 
$\tau$ power law limit; (iii) The power law index is related to the 
stretched exponential parameter $\alpha$ by Eq.  (\ref{conjecture2}); 
(iv) One can deduce the power law index from the small $\tau$ 
stretched exponential fit, even in situations where there is no 
information about the long tail in the data recorded. This is 
important for the main purpose of this paper. In fact, in Sections 
\ref{EEG} and \ref{music} we shall prove that EEG's and complex 
sounds, respectively,  are  examples of such a process because they 
are renewal processes and are well fitted by a stretched exponential. 
By means of Eq. (\ref{conjecture2}) we shall derive the hidden 
information on $\mu$.

The Mittag-Leffler (ML) function \cite{ml} is attracting an ever increasing
interest in the literature of complex fluids such as liquid crystals,
glass-forming liquids, and polymeric and colloid systems \cite{dejardin}. A
remarkable property of the ML function is that the complex susceptibility
produced by the ML relaxation function yields the  Cole-Cole
experimental form \cite{weron}. For this interesting property the reader can
consult also Refs.~\cite{dejardin},~\cite{ralf} and~\cite{coffey}.

The ML function establishes a bridge between the
stretched exponential behavior for short time and an inverse power law in the
long-time limit \cite{report}. This property is important, since it may settle
the controversy between the advocates of stretched exponentials and  those of
inverse power laws.
In the case of the financial market, the authors
of Ref.~\cite{raberto} have found that the ML function affords a very good
fitting of experimental data. There is a problem with the tails, however,
insofar as the lack of sufficiently rich statistics
make noisy the
time region where
the ML fat tail should appear \cite{bianco}. The
authors of
Ref. \cite{bianco} used the aging experiment to make an inverse power
law behavior distinctly emerge from the noisy background of the long time
regime.

Here we show an approach to the ML proving that in some noisy 
conditions only the stretched exponential portion of the ML function 
can remain visible. Our theoretical  approach refers 
to the time distance between two consecutive events, 
rather than a molecular relaxation process~\cite{weron}: The motor 
driving our process is a physical generator of events.
Let us assume 
that a natural time scale exists, where this physical generator of 
events is a Poisson process with rate
$r \ll 1$, so as to make the discrete time representation virtually 
indistinguishable from the continuous time picture. The time distance 
between two consecutive events produced by
this generator
is given by the waiting time distribution density
$\psi_{P}(n)$, which has the following exponential
form
\begin{equation}
\label{poisson}
\psi_{P}(n) = r exp(- r n).
\end{equation}
To generate the sequence of events of interest for
this paper, we operate as follows. We record the activity of this
physical generator at each and every time step of its natural time
$n$. At the same time, following the prescription of the
subordination theory~\cite{arne}, we turn the time $n$ into the continuous time
$t(n)$, by setting
\begin{equation}
t(n+1) - t(n) = \tau_{n}.
\end{equation}
It is straightforward to prove (see Appendix~\ref{appendix})
that the adoption of an inverse power law with no truncation and 
power index $\mu$ as subordination
function, together with the renewal assumption for the events production and
the condition $r \ll 1$, creates a  SP with the form of a ML function of order $\alpha$, denoted
by the symbol $\Psi_{SP}$, whose analytical expression is 
\begin{equation}
   \label{MLF}
   \Psi_{SP}(t) \equiv E_\alpha(-(\gamma t)^\alpha) = 
\sum_{n=0}^\infty (-1)^n \frac{(\gamma
   t)^{\alpha n}}{\Gamma(\alpha n + 1)},
\end{equation}
with
\begin{equation}
\label{conjecture2}
\alpha = \mu -1.
\end{equation}
In the case $0 < \alpha < 1$ it is well known that the ML function 
admits two limiting conditions, namely
\begin{equation}
   E_\alpha(-(\gamma t)^\alpha) \sim \frac{1}{(\gamma t)^\alpha} \qquad t \to
   \infty, \quad t > \frac{1}{\gamma}
\end{equation}
that is an inverse power law, and
\begin{equation}
   \label{stret_SP}
   E_\alpha(-(\gamma t)^\alpha) \sim exp(-(\gamma t)^\alpha) \qquad t \to 0,
   \quad t<\frac{1}{\gamma},
\end{equation}
which is a stretched exponential.\\
\indent To take into account that the data under study are finite, it is
convenient~\cite{arne} to select for the subordination function $\psi(\tau)$
an inverse power law with index $\mu$, which is exponentially truncated at $t
\gtrsim 1/\Gamma$, where $\Gamma < \gamma$. As an effect of this 
choice (see Appendix~\ref{appendix}) we get
for the function $\Psi_{SP}(t)$ a resulting form virtually indistinguishable
from a stretched exponential, in the intermediate time scale where 
the departure from the exponential form is significant. This derivation of the
stretched exponential is different from the one 
recently proposed by the authors of Ref. \cite{refereeA}. However, a 
comparison between our approach and that of these authors is not 
quite appropriate: in fact, Magdziarz and Weron \cite{refereeA} aim 
at the same purpose as that of the earlier work of Ref. \cite{weron}, 
the explanation of the non-exponential Cole-Cole relaxation, and do 
not afford prescriptions to evaluate the distribution of the time 
distances among consecutive renewal events, which is the main trust 
of this Section. 

The subordination process is realized by means of 
a random
prescription and, consequently, $\Psi_{SP}(t)$ is expected to fit the renewal
condition. We shall assess this property by means of the aging experiment of
Section~\ref{EEG}. We shall also establish a connection between the stretched
exponential $\exp (-(\gamma t)^\alpha)$, with $\alpha < 1$ and the power index
$\mu$ by means of the relation of Eq.~\eqref{conjecture2}
generated by the subordination procedure. Since the 
subordination is realized with a function $\psi(\tau)$, which is a 
truncated inverse power law, the resulting process in 
the long-time scale does not violate the ergodic condition. Thus  
the Poisson behavior, and the consequent lack of aging, as 
we shall see by means of the statistical analysis of both  EEG data 
and music composition, is recovered. The same limitations are shared by many other 
complex processes, see for example \cite{arne}, due to the obvious 
fact that an exact inverse power law behavior is an idealization that 
would imply the infinite size of the systems under study. The study of this
idealized condition is useful to shed light into the transient behavior before
the eventual Poisson condition.

\section{The minimum spanning tree approach as a generator of events}
\label{MST}
\noindent In this Section we introduce an algorithmic procedure that
will allow us to process the data at our disposal so as to generate 
events and consequently
time series to analyze. The method is based on the
famous Minimal Spanning Tree (MST) algorithm~\cite{book}, which we now briefly
introduce.\\
\indent To define the MST approach we closely follow the arguments of
Ref.~\cite{StanleyMantegna}. Imagine a data set consisting of $N$ 
columns, each column
representing the signal recorded by an electrode, or the timber, 
pitch, note \ldots, etc, of the music
composition. Taking two columns, say $\bm x$ and $\bm y$, we define 
the correlation
coefficient between the two columns as follows
\begin{equation}
\label{correlation}
\rho_{\bm{xy}} = \frac{\sum_{k=0}^t (x_k - <{\bm x}>)(y_k - <{\bm
     y}>)}{\sigma_{\bm x} \sigma_{\bm y}},
\end{equation}
where the quantities $< \bm x>$ and $\sigma_{\bm x}$ are, 
respectively, the average and
the standard deviation of the values that $\bm x$ takes over the interval
$[0,t]$. Consider now the following quantity
\begin{equation}
   \tilde{\bm x} = \frac{{\bm x} - <{\bm x}>}{\sigma_{\bm x}}
\end{equation}
and assume that in the interval $[0,t]$ there are $t$ values of the vector
$\tilde{\bm x}$, namely $\tilde{x}_k, k = 1,\ldots, t$. The distance 
between two
columns $\tilde{\bm x}$ and $\tilde{\bm y}$ over the time interval $t$ 
is easily obtained adopting the well known
formula of the Euclidean distance
\begin{equation}
   \label{distance_squared}
   d_{\bm {xy}}^2 = ||\tilde{\bm x} - \tilde{\bm y}|| = \sum_{k = 1}^t 
(\tilde{x}_k - \tilde{y}_k)^2.
\end{equation}
Moreover $\sum_{k=1}^t \tilde{x}_k^2 = 1$. Therefore, Eq.~\eqref{distance_squared} becomes
\begin{equation}
   \label{distance_squared2}
   d_{\bm {xy}}^2 = \sum_{k=1}^t (\tilde{x}_k^2 + \tilde{y}_k^2 - 2\tilde{x}_k
   \tilde{y}_k) = 2 - 2 \sum_{k=1}^t \tilde{x}_k \tilde{y}_k.
\end{equation}
The last sum on the right hand side of Eq.~\eqref{distance_squared2} is the
correlation coefficient between $\bm x$ and $\bm y$ of 
Eq.~\eqref{correlation} at the time $t$. It is
therefore possible to define the distance between the columns $\bm x$ and $\bm
y$ as
\begin{equation}
d_{\bm {xy}} = \sqrt{2(1-\rho_{\bm{xy}})}.
\end{equation}
Since $-1 < \rho_{\bm{xy}} < 1$, then $0 < d_{\bm {xy}} < 2$. Moreover
$d_{\bm {xy}}$ fulfills the property of a distance~\cite{StanleyMantegna}.\\
\indent Through the distance $d_{\bm{xy}}$ we now introduce the MST
approach over the time interval $t$. In a connected graph of $N$ 
objects, weighted through
the distances $d_{\bm{xy}}$, the MST is the tree with $N-1$ links for which the
total sum of the edges is minimum. In the literature there are many 
algorithms to create the MST. Here we select the method proposed by
Kruskal~\cite{kruskal}. This method consists of the following steps: we sort
the distances in increasing order; we select the shortest distance and draw an
edge between the associated nodes; we go to the next distance and draw edges;
if an edge creates a loop we erase it; we continue drawing lines until all the
$N$ columns are represented. \\
\indent In this paper we  make a  dynamical use of the
MST. We build the MST over a time windows of length $t$, then we move to the
next (non overlapped) window and build the MST corresponding to the 
new position. 
An event is defined whenever a change in the distribution of links occurs. This way of
proceeding allows us to define events and consequently the distribution of
waiting times between two consecutive events. Moreover, when building the distribution
of the number of edges, we do not label our electrodes, but simply 
look at the shape of the distribution, that essentially characterizes the
topology of the MST. If distributions at subsequent times do not exactly
coincide, we consider it to be a signal of a change in the global
properties of the brain, and indicate this as our critical event. \\

\indent The MST has been recently adopted in the literature in an increasing
number of papers on different complex systems, from networks
(e.g.~\cite{stanley} among the others) and financial markets 
(e.g.~\cite{mantegna} among the
others) to neuro-physiological processes~\cite{epilepsy}.
Particularly, we quote
the recent paper of McDonald \emph{et al.}~\cite{johnson}.
In this paper the MST approach has already been used and applied to financial
time series, but in a  different way. In fact, the authors of
Ref.~\cite{johnson} monitor the survival-ratio of edges in time, that is, the
ratio between the number of edges connected to a certain node at time
$0$ and the
same quantity computed at a later time $t$, and find that it decays
non exponentially.\\

 According to Ref.~\cite{johnson}, the MST, by definition, is more suitable to study
positively  than negatively correlated systems. Indeed, negatively correlated electrodes
contribute differently to the MST than positively correlated electrodes. In our
analysis this criticism only relatively applies, as we are interested in the
changes of the MST topology in time, i. e. in the changes of the correlations with time. 

\indent In the next Section we shall analyze the statistical properties of
the time series obtained in the aforementioned way.

\section{Analysis of EEG}\label{EEG}
Digital EEG data was collected on $5$ subjects, utilizing a Scan LT-40 amplifier, manufactured
by Neuro Scan Medical Systems, El Paso, Texas. The Scan LT-40 is an FDA
approved medical device for the collection of digital EEG. Online monitoring
of EEG was provided by NeuroScan Medical Systems 1.2 software.  Offline
evaluation and the removal of artifacts from the EEG record and the conversion
of the data to ASCII files were accomplished using NeuroGuide 2.2.6 software.
An electrode cap from Electrocap International Inc. was used to provide
standardized electrode placement. Digital EEG data was collected from 
19 locations using the international 10-20
system of electrode placement.  A reference electrode was placed on each
earlobe to provide a linked ears montage for the physical reference of the
scalp recordings.  The impedance of the respective earlobe reference
electrodes was maintained within 1 $K\Omega$ of each other. All other electrode
impedances were maintained at 1 or 2 $K\Omega$ relative to amplifier input
impedance with no more than 1 $K\Omega$ of variance between any of the electrode
contacts.   The amplifiers used to acquire the EEG were calibrated with sine
waves before the acquisition of EEG.  The EEG data was digitized at a rate of
250 samples per second. Before Analog to Digital conversion, anti-aliasing was
achieved by a low-pass filter built into the software.  The EEG was visually inspected online during
acquisition to monitor for artifact. When necessary, data collection was
stopped to identify and remove persistent sources of artifact such as muscle
tension.  The subjects included $2$ healthy subjects, $1$
subject with chronic back pain, and $2$ subjects with mental
depression. Subjects were  medication free during the data
acquisition. The average number of data collected is $26,141$, corresponding
to about $104$ seconds of record. The minimum number of data for a single
subject is $10,961$, corresponding to about $43$ seconds of recording, for one
of the healthy subjects, while the maximum number of data for a single subject
is $56,544$, corresponding to about $226$ seconds, for the subject with
chronic back pain. Only two states were considered for the subjects, namely
eyes open (EO) and eyes closed (EC). The acquisitions were made in the same
conditions for all the subjects.\\

\begin{figure}[!ht]
  \includegraphics[angle=270,scale=0.5]{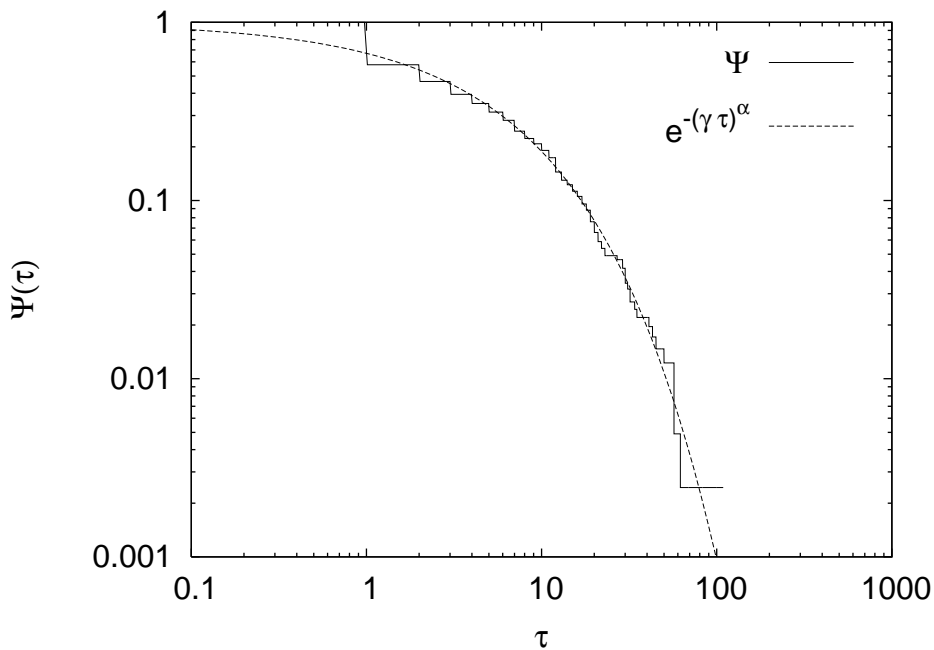}
  \caption{The SP of MST topological changes for the individual with back pain
    in the EO condition. The curve is fitted using  a
    stretched exponential of the form of Eq.~\eqref{stret_SP} with $\gamma =
    0.205$ and $\alpha = 0.595$.}\label{distr}
\end{figure}

\indent The MST approach, discussed in the previous Section, is applied to the set of 19
columns that the data acquisition method affords.  The time series of MST
topological changes is obtained and, to be consistent with the theoretical
results of Section~\ref{stretched}, the SP $\Psi(t)$ of the time distances
between two consecutive MST topological changes is evaluated. In Fig.~\ref{distr} we show the
SP relative to the individual with back pain in EO condition.
We fit the resulting curve with a stretched exponential of the form of Eq.~\eqref{stret_SP} with
parameters $\gamma = 0.205$ and $\alpha = 0.595$.

Similar results have been obtained for the SPs of all the other individuals at our disposal.
Using Eq.~\eqref{conjecture2}, we find that the values of $\mu$ for the EO condition lie between $1.595 \pm 0.005$ and
$1.960 \pm 0.013$, while for the EC condition between $1.746 \pm 0.018$ and
$1.974 \pm 0.024$. \\ 

\begin{figure}[!ht]
  \includegraphics[angle=270,scale=0.5]{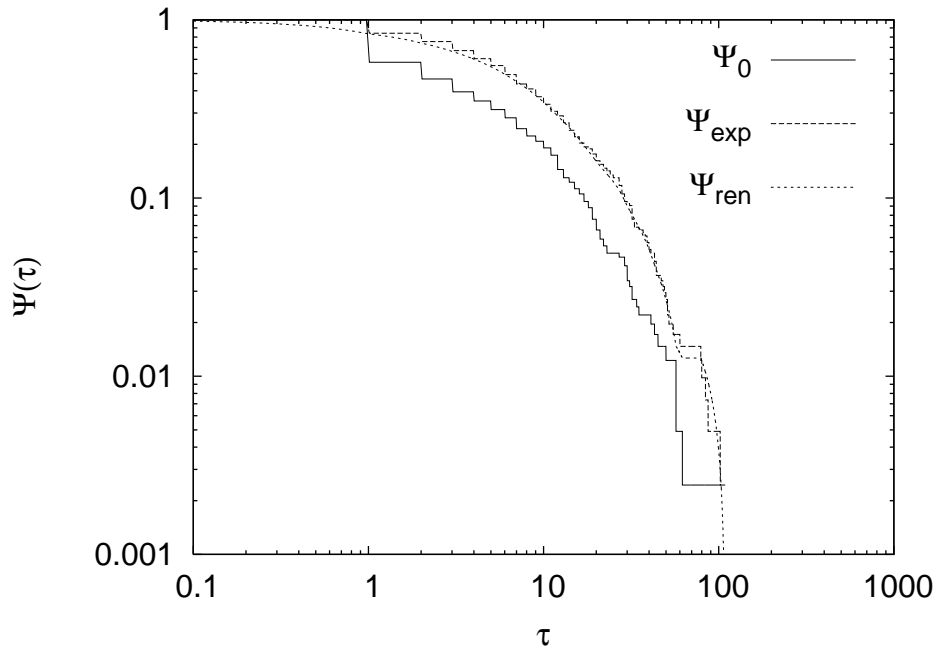}
  \caption{The AE analysis on the time series of Fig.~\ref{distr}, for $t_a = 30$. The aging
    is present and compatible with the renewal condition. }\label{aging}
\end{figure}

The results of Fig.~\ref{distr} suggest that the  process is  strongly non-Poissonian. However, the mere
analysis of the distribution of waiting times is not enough to establish the
real nature of the process. In fact, as pointed out in
Refs.~\cite{paradiso, cpl}, a modulated Poisson process as well can have as
an outcome a non-exponential waiting time distribution. In order to assess if the time series
under study is produced by a genuinely non-Poisson renewal process, the authors
of Ref.~\cite{paradiso} propose the Renewal Aging Experiment (AE).
According to this procedure, in addition to the waiting time density
$\psi(\tau)$,  an aged waiting time density $\psi_{exp}(\tau, t_{a})$ 
must be evaluated. This is done as
follows. Let $\{t_{i}\}$ be the series of time obtained through the
prescription indicated in the previous Section. For each time $t_{i}$
the first time of the sequence at a distance from $t_{i}$ equal to or 
larger than $t_{i} +
t_{a}$ is recorded. This time will be $t_{k}$, with $k > i$. The time distance
$\tau(t_{i}, t_a) = t_{k} -(t_{i} + t_{a})$ is considered. The procedure is
repeated for all the times of the sequence $\{t_{i}\}$, and the 
sequence of these recorded
time distances is used to generate the distribution density 
$\psi_{exp}(\tau,t_{a})$. Moreover, the
following quantity is evaluated
\begin{equation}
\psi_{ren}(\tau, t_{a}) = \frac{\int_{0}^{t_a} dy 
\psi_{num}(\tau+y)}{K(t_{a})},
\end{equation}
where $K(t_{a})$ is a suitable normalization constant and $\psi_{num}(t)$ is
the numerical waiting time distribution corresponding to $t_{a} = 0$.\\
\indent To establish the renewal character of the process by means of the AE
it is convenient to plot the corresponding SP's, $\Psi_{ren}(\tau, t_a)$ and
$\Psi_{exp}(\tau, t_a)$. The SP of age $t_a = 0$, indicated as $\Psi_0$, is
also plotted. In Fig.~\ref{aging} the three SPs are compared, and it
is found that $\Psi_{exp}$ and $\Psi_{ren}$ virtually coincide. According to Ref.~\cite{paradiso} this
good accordance is the numerical evidence of the renewal nature of the process.

Moreover this result confirms the theoretical
framework of Section~\ref{stretched}, allowing us to obtain the index of the
inverse power law distribution  from the index of the
stretched exponential, namely from Eq.~\eqref{conjecture2}.
Since all the values of $\alpha$ are smaller than $1$ we conclude that $\mu <
2$, and therefore  that the human brain is a non-ergodic system. It 
is important to repeat here what we already pointed out in Section 
II, that in the long run, as all the anomalous processes occurring in 
nature, the ergodic properties, as well as the Poisson condition and 
the associated lack of aging, are recovered. \\

How can we explain the emergence of this stretched exponential of renewal origin?
An unknown referee pointed to us that there may exist a dominant 
structure and that the spanning trees appearing in the processed 
data may produce short and long waiting times according to their 
distance from the dominant structure. If there is a systematic 
alternation of short and long times, we end up in a model recently 
discussed in Ref.~\cite{cpl}.  In his case there would be a 
drastic reduction of renewal aging. If on the contrary the process is 
not a systematic alternation of short and long waiting times, it 
becomes indistinguishable from the non-Poisson model that we propose 
with the theory of Section~\ref{stretched}.\\

Now we have to address an important issue, concerning the 
physical origin of the complexity property revealed by our analysis. 
We share the opinion of Haken \cite{haken} that the brain global 
behavior is an emergent property produced by a synergetic  process of 
synchronization \cite{barkailikesit}. As a consequence the different 
electrodes are correlated and the MST of Section III detects the 
emergence of this global property. The electrodes on which the MST 
analysis is based, are single units, like the interacting columns of 
a surface growing as an effect of random deposition \cite{arne}. Due 
to the cooperation with the other columns the single columns inherit 
the complexity of the whole growing surface \cite{arne}. \\

On the 
basis of this observation, we are tempted to make the conjecture that 
the single electrodes inherit the global complexity. It is important 
to stress that Buiatti and his co-workers \cite{marcobuiatti} have 
recently obtained results confirming this conjecture, with a method of 
analysis based on the observation of a single EEG. Our conjecture is 
also compatible with the results found by the
authors of Ref. \cite{fbm}. To understand the connection between the results
of this section and those found by the authors of Ref. \cite{fbm} the reader
should consult the recent article of Ref. \cite{rasit}. The authors of this
paper have studied the dynamic approach to Fractional Brownian Motion (FBM),
as expressed by
\begin{equation}
\label{mandelbrot}
\frac{d}{dt} x(t) = \xi(t),
\end{equation}
where $\xi(t)$ is a fluctuating velocity with memory, namely, a correlation
function with slow tails. The dynamical approach to FBM proves that the
variable $x$ in the asymptotic time regime shares the same properties as the
traditional form of FBM. They have also proved that the origin recrossing of
$x(t)$ generates a non-Poisson renewal process with the power
$\mu$ related to the FBM scaling coefficient $H$ by
\begin{equation}
\label{fromhtomu}
\mu = 2 - H.
\end{equation}
Note that the adoption of the more realistic model of 
the interacting columns of an interface growing as an effect of 
random deposition of particles would produce saturation and also a 
truncation of the inverse power law waiting time distribution, a 
property similar to the long-time Poisson behavior revealed by Fig. 1 
and Fig. 2. However, if the observation is limited to the short-time 
region, the condition of Eq. (\ref{fromhtomu}) applies.

The analysis of the authors of Ref. \cite{fbm} rests on $\xi(t)$, rather than
on $x(t)$. We are convinced in fact that the single electrodes yield a signal
that has to be interpreted as the variable $x$ of Eq. (\ref{mandelbrot}). Thus
to relate the results of Ref. \cite{fbm} to the results of this Section, we
must adopt Eq. (\ref{fromhtomu}). This procedure generates $\mu < 2$. With
this interpretation in mind, we find that also the analysis of Ref.~\cite{fbm}
proves that individuals whose signal was recorded in the EC condition is
closer to $\mu = 2$ than the individuals in the EO condition.

Stressing this result is a way for us to draw the
attention of the reader on the fact that the discovery of non-Poisson renewal
events, made possible by the method adopted in this paper, does not yield
results conflicting with the work of other groups. In Table~\ref{mark}, we
present a sample of the  results of our analysis on our group of
individuals. We remind that two of them were healthy, one had back pain, and
two had mental depression. The values of $\mu$ refer to the
same subject, in the two conditions. We see that the parameters $\mu$ of 
individuals in the EC condition are significantly closer to $2$ than the parameters $\mu$ of
patients in the EO condition.
In Table~\ref{cinesi} we report a sample of the results
of Ref. \cite{fbm} (healthy individuals only, Tab. I) in terms of $\mu$, by means of the rule of
Eq. (\ref{fromhtomu}). By comparing the results of Table~\ref{mark} with
those of Table~\ref{cinesi}  we reach the conclusion that the results
of our analysis are compatible with those of Ref. \cite{fbm}. It is necessary
to stress that this result is not statistically significant, given the small
size of our sample, but it represents indeed a trend, compatible with the
the results present in the literature. More research work is necessary to be
done to confirm the presence of this interesting effect. 

\begin{table}[!ht]
    \begin{tabular}{c|c|c}
      \multicolumn{3}{c}{$\mu$}\\
      \multicolumn{3}{c}{}\\
      \hline
      Id & EO & EC \\
      \hline
      H1 & 1.730 (0.010) & 1.800 (0.010) \\
      H2  & 1.740 (0.012)& 1.770 (0.013)\\
      BP & 1.595 (0.005) & 1.746 (0.018) \\
      D1 & 1.748 (0.013) & 1.787 (0.014)\\
      D2 & 1.960 (0.023) & 1.974 (0.024)\\
    \end{tabular}
\caption{This Table shows a sample of the results of the analysis on our group of 
      individuals in the EO (left side of the Table) and EC (right
      side of the Table) conditions. Results show a larger value of $\mu$ for 
 EC condition. In parenthesis we report the standard deviation. The symbol H
      indicates a healthy individual, BP the individual with back pain, D the
      subjects with mental depression.}\label{mark} 
\end{table}

\begin{table}[!ht]
    \begin{tabular}{c|c}
      \multicolumn{2}{c}{$\mu$}\\
      \multicolumn{2}{c}{}\\
      \hline
      EO & EC \\
      \hline
      1.261 & 1.888 \\
      1.242 & 1.931 \\ 
      1.294 & 1.898 \\
      1.278 & 1.877 
    \end{tabular}
    \caption{This Table shows a sample of the results of the analysis of
      Ref.~\cite{fbm} on healthy individuals in the EO (left side of the
      Table) and EC (right side of the Table) conditions. The parameter $\mu$ has been obtained according to
      Eq.~\eqref{fromhtomu}. Results show a  larger value of
$\mu$ for EC condition.}\label{cinesi}
\end{table}

Before ending this Section, we want to stress that the main result 
of this paper, namely the surprising complexity matching between 
brain and complex sounds that we shall discuss in Section~\ref{music}, does 
not depend in any way on the conjecture we make about the emergence 
of $\mu < 2$ in the single electrodes. This conjecture would have the 
effect of explaining the findings of Refs. \cite{marcobuiatti,fbm}, 
but, if proved wrong, would not weaken the validity of the main 
result of this paper.

\section{Analysis of music composition}\label{music}
The music composition data that we analyze in this Section have been produced
by means of a virtual instrument (software synthesizer) suitably 
designed \cite{winsor1,winsor2} to produce
physiological effects \cite{winsor3,winsor4}. The instrument is capable of
generating abstract sonic textures that are free from overt cultural
influences. Moreover, through the use of Presets, a record can be kept of all
parameters of  consequence in the generation of the musical material. This
feature is important for purposes of correlating the output sonic textures
with, for instance, biological data from EEG recordings. In this way,
researchers can precisely pinpoint areas of interest in biological data for
analysis of and comparison with the generative parameters of the sound
structures. Designed to allow control over the redundancy of
time-point and pitch/frequency patterns in a hierarchical framework, the
virtual instrument sonic textures can be gradually morphed between constant
states (stable) regimes and chaotic (complex) regimes via controls built into
the graphical user interface. Moreover, precise measurement and recording of
all generative schemes is possible, as well as synchronization with the EEG
data time series. An important characteristic of the instrument is that
vertical sonic textures are flexibly configurable with respect to the degree
of vertical and horizontal redundancy of pitch- and time-space organization
within the ongoing sonic flow. Various degrees of complexity can be
introduced via Presets prior to and during the transmission of audio 
data to the test
subject.  For instance, microtonal controls are built into the instrument, so
that the researcher can regulate the content of the sonic texture at any given
point to create a relative harmonicity shift  of the composite sonic data
stream.  In other words, there is maximum control over the degree of
complexity via overlapping, phase-modulated patterning of melodic, harmonic,
rhythmic, and texture-density data.

\begin{figure}[!ht]
  \includegraphics[angle=270, scale = 0.5]{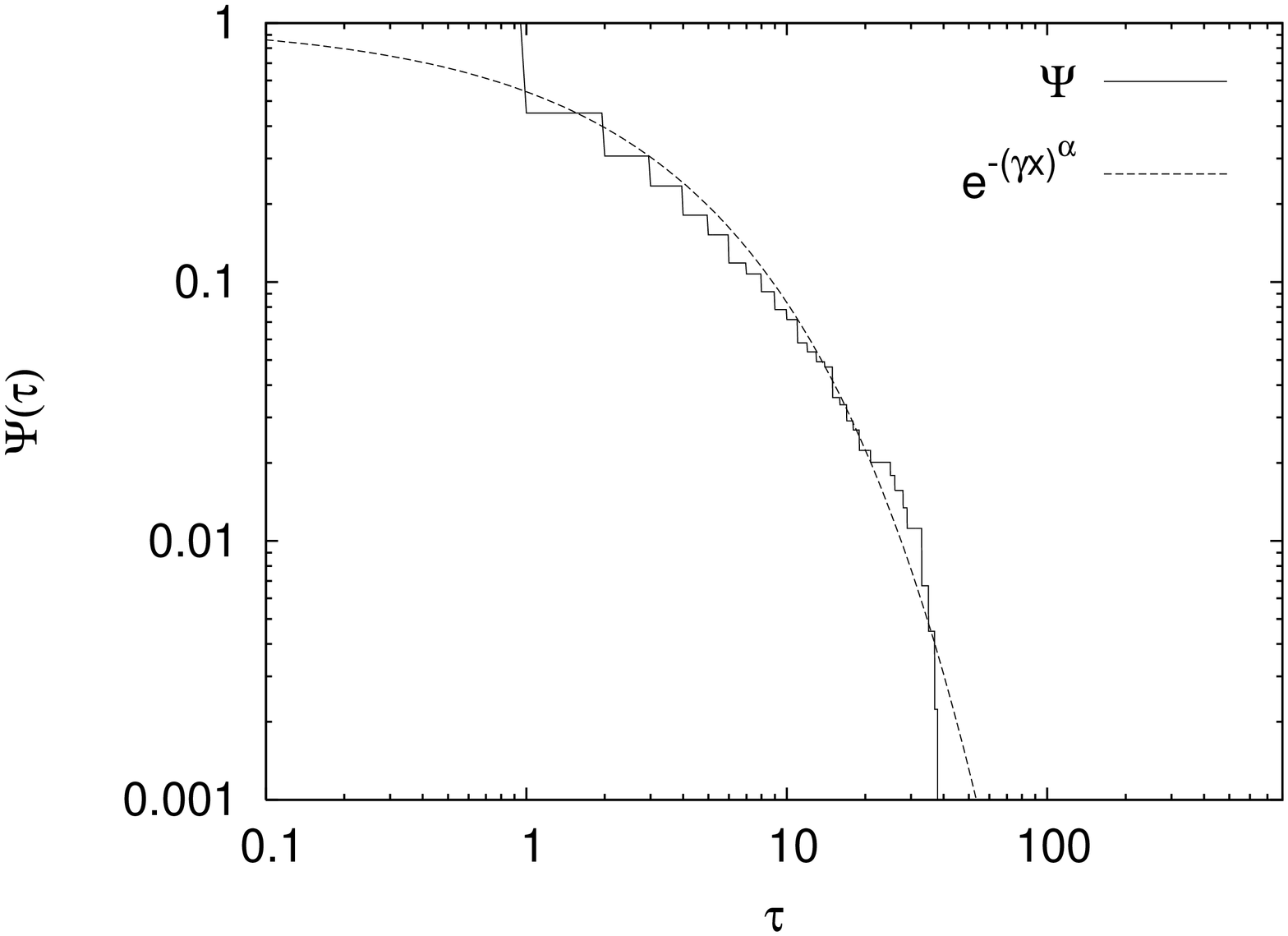}
  \caption{The SP of MST topological changes obtained from the music
    composition. The best fitting function is a stretched exponential of the
    form of Eq.~\eqref{stret_SP}, with $\gamma = 0.430$ and $\alpha =
0.600$. $t_a = 100$.}\label{distr_music}
\end{figure}

The instrument produce a vector $ V(t)$, with $7$ components, namely pitch
onset time (in milliseconds), note frequency (in Hertz), waveform type,
amplitude, articulation, preset number, and oscillator number. All
these components are filed and assumed to afford information about the signal.
We therefore adopt for the music composition the same procedure
applied before for the
EEG signal, building also in this case a time series of MST topological
changes. The resulting SP for a sample record is plotted in Fig.~\ref{distr_music}.

\begin{figure}[!ht]
  \includegraphics[angle=270, scale = 0.5]{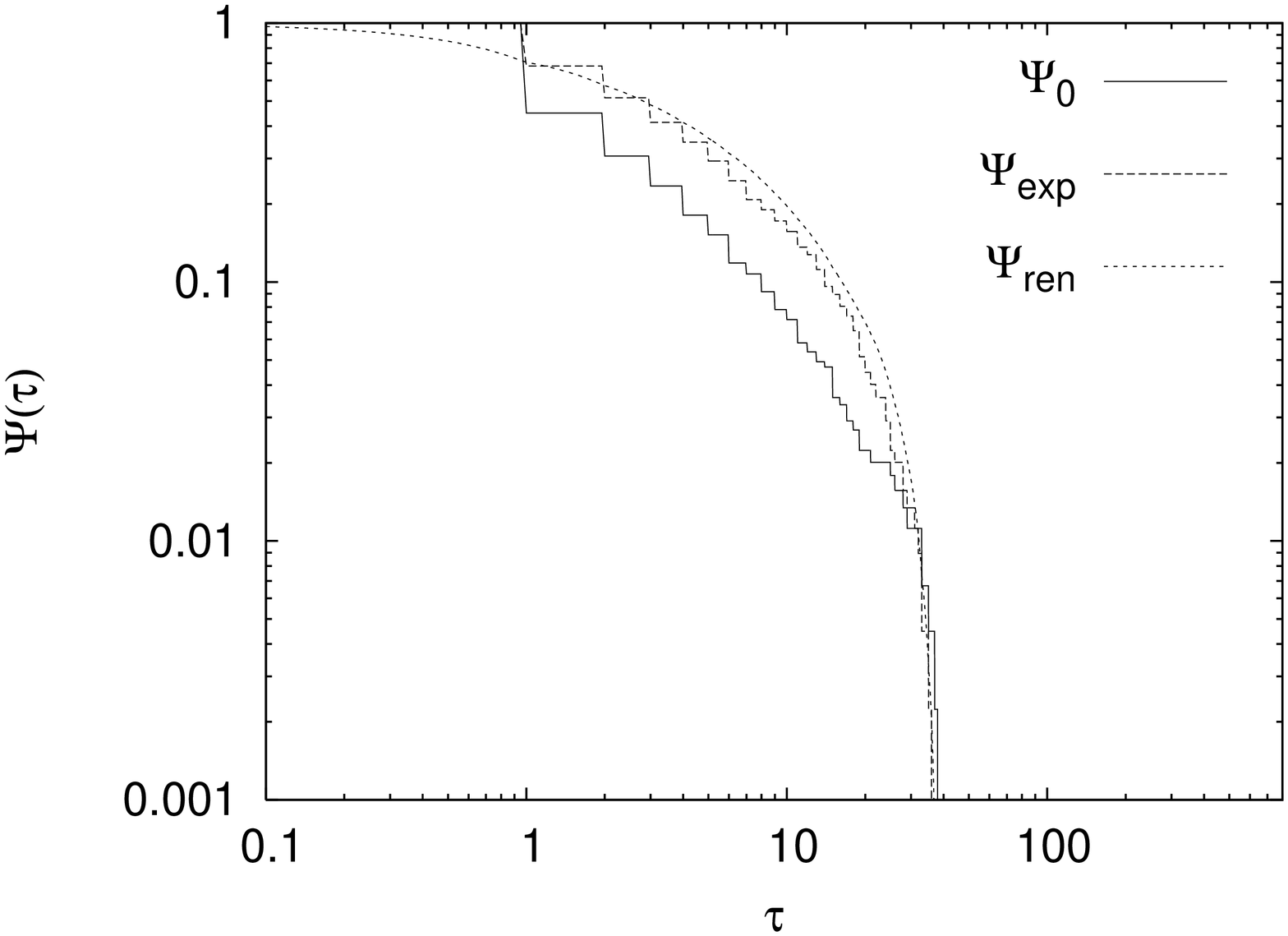}
  \caption{The AE on the time series of Fig.~\ref{distr_music}. The aging
    effect is compatible with the renewal assumption.}\label{aging_music}
\end{figure}

We see from the figure that also in this case the distribution of
topological changes produces a non exponential SP, namely a stretched
exponential with parameters $\gamma = 0.430$ and $\alpha = 0.600$. As in
Section~\ref{EEG}, a further test is needed to prove that the process is
renewal. The results of the AE are plotted in Fig.~\ref{aging_music}.

Also in this case the AE supports the renewal assumption for the process.
We conclude that music composition shares the same properties as the EEG
signal analyzed in Section~\ref{EEG}. This suggests that both human brain and
music composition, sharing the condition $\mu < 2$, are complex systems that,
in the absence of the exponential truncation, would violate the ergodic condition~\cite{barkai}.

\section {A short review of the CM effect}\label{CM}
The authors of Refs.~\cite{barbi,gianluca} have developed a linear
response theory that applies to non-Poisson renewal systems. The form
of this linear response theory is given by:
   \begin{equation}
   \label{key}
\Pi(t) \equiv  \langle \xi_{S}(t) \rangle =  \epsilon \int_{0}^{t}
dt^{\prime} \chi(t,t^{\prime}) \xi_{P}(t^{\prime}),
   \end{equation}
   where $\xi_{S}(t)$ denotes the signal produced by the system of
interest, $\chi(t,t^{\prime})$ is the linear response function and 
$\xi_{P}(t)$ is the external perturbation.  In the case of the human 
brain the signal $\xi_{S}(t)$ is a
global property emerging from the synchronization of different brain
areas, roughly corresponding to the superposition of the signals
detected by the electrodes fixed on the patient's scalp. In the
absence of perturbation this global signal is characterized by the
non-Poisson renewal events revealed by the method described in
Section \ref{EEG}. To transform these data into $\xi_{S}(t)$, we
assign alternate signs to the quiescent time regions between two
consecutive events. The complexity of this signal is denoted by means
of the power index $\mu_S < 2$, emerging from the analysis of Section
\ref{EEG}.

  If the perturbation function $\xi_{p}(t)$ is harmonic the system does not
respond \cite{barbi}. This is a clear sign of complexity, insofar as
a non-Poisson renewal signal cannot be interpreted as the
superposition of infinitely many independent processes. The
individual constituents of a set of neurons responsible for any
cognitive action are expected to be strictly cooperating with all the
others' constituents. As a consequence, a harmonic perturbation
triggers a cascade of different time scales, thereby violating  the
prescriptions of ordinary stochastic resonance processes
\cite{barbi,gammaitoni}. Thus, we are led to make the conjecture that
the transmission of information from the perturbing signal to the human
brain is determined by the interaction between the renewal events of
the perturbing signal and the renewal events of the perturbed system.
If the perturbing signal does not have any renewal events, as in the
case of a harmonic perturbation, there is no response to a weak
perturbation.

More recently the authors of Ref.~\cite{letushope,cme}
have proved that this conjecture is correct, and  that in the case
where $\xi_{P}(t)$ is a signal derived from another non-Poisson
renewal system with index $\mu_{P} < 2$,
   the system responds, and the intensity of the response is maximum
when we use the matching condition $\mu_S = \mu_P$\cite{letushope}. 
The authors of
Ref. \cite{letushope,cme} have denoted this effect with the name of
Complexity Matching (CM), and they proved that when $\mu_P < \mu_S$ 
the perturbed system inherits the perturbation power index. In 
Section \ref{music} we have proved that
music composition is actually a complex signal with $\mu_P < 2$.
Thus, the existence of the CM effect leads us to conjecture that the
reason why the brain is sensitive to music lies on the fact that both
the brain and music are non-Poisson renewal systems living in the
non-ergodic region.

\section {Concluding Remarks}\label{theend}

The literature on
complexity is wide, and there are many different proposals to account
for complex processes. All these proposals share only one essential
property, this being the departure from the canonical exponential
distribution of ordinary statistical physics. The authors of Ref.
\cite{paradiso} have pointed out that this departure can be realized
by means of quite different physical processes, either
non-homogeneous Poisson processes or  homogeneous non-Poisson
processes. The different physical origin of these two processes is
revealed by the AE.

According to the authors of Refs.
\cite{wrong1,wrong2} there exists a close connection between
self-organized criticality \cite{bak}, superstatistics \cite{beck}
and non-extensive thermodynamics \cite{tsallis}. This analysis
coincides with the critical view illustrated by Jensen \cite{jensen}.
It is very attractive to conjecture that the brain operates at or
near a self-organized critical state \cite{wrong3,wrong2}. This
corresponds to the recent observation \cite{katja,barkailikesit}
that neuron synchronization  is a sort of phase transition involving
a close  cooperation among the elementary constituents of the neuron
set.
  However, in our opinion, the main limits of these interesting
theories is that they do not pay attention to the important role of
renewal events, whose objective existence is made compelling by the
results of the AE, in both the case of blinking quantum dots
\cite{paradiso} and of neuron synchronization \cite{barkailikesit}.

We think that, although these properties have been overlooked by the
majority of the researchers working in the field of complexity, they
deserve more attention, and we hope that this paper may serve the
important purpose of raising the interest of the investigators in
this direction.

  If the importance of these non-Poisson renewal
events is not recognized, the results of the analysis of this paper
on both EEG data and music composition, which are based on a solid
method of statistical analysis \cite{paradiso}, are incomprehensible.
If, on the contrary, we accept the leading idea that complexity rests
on the close cooperation of elementary components losing their own
identity at the moment of the onset of synchronization, then the fact that
both music composition and the human brain are non-Poisson renewal
processes becomes a natural way of explaining why music exerts its
influence on the brain. 

This conclusion is so important as to deserve 
further remarks. First of all, a more appropriate term to denote the 
music composition analyzed in this paper would be \emph{complex 
sound}. In fact, the main purpose of the music composition utilized in this
work is to affect brain complexity rather than generate emotional and/or
aesthetic responses, per se.
Thus, to make the composition of these complex sounds more flexible and more suitable 
to the purpose of realizing an efficient transport of information 
from acoustic excitation to the human brain, thanks to the CM effect 
\cite{letushope,cme}, we have to focus  on the realization of a given 
$\mu_{P}$ more than on esthetic purposes. On the basis of the CM 
effect \cite{letushope,cme} the network of EEG electrodes is expected 
to inherit the same $\mu_{P}$ as that of the acoustic excitation if 
$\mu_{P} < \mu_{S}$, where $\mu_{S}$ denotes the brain complexity 
index. We plan to design a real experiment based on recording EEG 
data from a subject listening to music with an exponent $\mu_P$.We 
plan also to  observe whether the exponent of the EEG changes if the 
listeners hears songs with different exponents. All of this requires 
much more work, and the solution of technical as well as conceptual 
problems. However, we are convinced that the results of this paper 
are important enough as to trigger further research work for the 
realization of this important experiment.

\appendix
\section{}
\label{appendix}
Let us address the problem of establishing the SP $\Psi_{SP}(t)$
corresponding to a set of sequences $\xi(t)$ prepared at $t= 0$. It
is evident that
\begin{equation}
\label{explain}
\Psi_{SP}(t) = \sum_{n=0}^{\infty} \int_{0}^{t} dt^{\prime}
\psi_{n}(t^{\prime}) \Psi(t-t^{\prime}) exp(-rn).
\end{equation}
To explain Eq. (\ref{explain}), let us notice that
$\Psi_{SP}(t)$ is 
the probability that the physical generator does not produce any 
further event, after the initial preparation event, until time $t$.
In the natural time scale the probability density of not producing an 
event is $exp(-rn)$. To evaluate $\Psi_{SP}(t)$, we have to multiply 
$exp(-rn)$
 by
$\psi_{n}(t^{\prime})$ and by $\Psi(t-t^{\prime})$. This is because
$exp(-rn)$ indicates that until time $t$ the physical generator acts 
$n$ times, thereby making $n$ drawings from the distribution density
$\psi(\tau)$. Note that  $\psi(t)$ is also called  subordinating 
function \cite{arne}. This function is the probability distribution 
density affording information on the time distance between two 
consecutive actions of the physical generator, not necessarily 
producing events. With no drawing ($n=0$), we get $\Psi_{SP}(t) = 
\Psi(t)$, where $\Psi(t)$ is the SP corresponding to $\psi(t)$, 
namely, the probability that the physical generator does not act 
until time $t$. With
$n$ drawings we
fill a time interval of length $\tau_1 + \tau_2 +\cdot \cdot \cdot \tau_n =
t^{\prime} < t$. The factor $\Psi(t-t^{\prime})$ ensures that no drawing
occurs in between $t^{\prime}$ and $t$. The function $\psi_{n}(t^{\prime})$
denotes the probability that $n$ drawings from the distribution $\psi(\tau)$
occurred, the last of which occurred exactly at time $t^{\prime}$. 
Due to the renewal nature of this process
we have
\begin{equation}
\psi_{n}(t) = \psi_{n-1}(t) \otimes \psi_{1}(t),
\end{equation}
where $\otimes$ indicates time convolution
and $\psi_{1}(t) = \psi(t)$.

The Laplace transform of $\Psi_{SP}(t)$, $\hat\Psi_{SP}(u)$, is
expressed as a function of the Laplace transform of $\psi(t)$,
$\hat\psi(u)$, as follows
\begin{equation}
\label{exact}
\hat \Psi(u) = \frac{1}{\left[1 - \hat \psi(u) exp(-r)\right]}
\frac{1}{u}\left[1 - \hat \psi(u)\right].
\end{equation}
To obtain the previous expression, still no hypothesis on the form of
$\psi(\tau)$ has been done. If $\psi(\tau)$ is an exponential, the 
subordination exerts no
physical effect on the SP, as its form remains exponential. Let us see the
case of non exponential subordination function. With straightforward algebra,
and assuming $r \ll 1$,
Eq.~\eqref{exact} becomes:
\begin{equation}
   \hat{\Psi}_{SP}(u) = \frac{1}{u + r \hat{\Phi}(u)} \qquad \hat{\Phi}(u) =
   \frac{u \hat{\psi}(u)}{1 - \hat{\psi}(u)}
\end{equation}
that is the Laplace transform of
\begin{equation}
 \label{convolution}
   \frac{d}{dt}\Psi_{SP}(t) = -r \int_0^t \Psi_{SP} (t-t') \Phi(t') dt'
\end{equation}
and $\Phi(t')$ is a memory kernel. Note that the assumption $r \ll 1$ is 
not necessary to generate the time convoluted structure of Eq. 
(\ref{convolution}). It is essentially required to make 
 the natural 
time $n$ compatible with a continuous time representation, thereby 
yielding for the survival probability in the natural time scale the 
exponential form $exp(-rn)$. An even more consequence of $r \ll 1$ is 
that, as we shall see hereby, this condition makes very extended 
the stretched exponential regime.  

In the case where the memory kernel $\Phi(t)$ is a delta of Dirac, 
Eq. (\ref{convolution}) makes $\Psi_{SP}(t)$ become an ordinary 
exponential. 
 To generate a stretched exponential we must make a 
proper choice of the memory kernel $\Phi(t)$, and consequently 
of the subordination function. Let us assign to the memory kernel in 
the Laplace space the following form
\begin{equation}
   \label{memory_kernel}
   \hat{\Phi}(u) = \chi (u + \Gamma)^{2-\mu} \qquad \chi \equiv [\Gamma(2-\mu)
   T^{\mu -1}]^{-1}.
\end{equation}
and let us assume the parameter $\mu$ to be smaller than $2$.
With this choice the SP becomes, in the Laplace domain,
\begin{equation}
   \label{survival_ML}
   \hat{\Psi}_{SP}(u) = \frac{1}{u + \gamma^\alpha (u + \Gamma)^{1 - \alpha}}
\end{equation}
with
\begin{equation}
   \label{relation_mu_alpha}
   \alpha \equiv  \mu-1 \qquad \gamma \equiv (\chi r)^{\frac{1}{\alpha}}.
\end{equation}
If $\Gamma = 0$, we recognize in Eq.~\eqref{survival_ML} the well 
known Laplace transform of a
ML function of order $\alpha$, see Eq.~\eqref{MLF}.
For $\alpha = 1$, the ML function becomes an exponential. 
 
 Note 
that the parameter $\Gamma$ has the important role, as in Ref. 
\cite{arne}, of taking into account that we are working with systems 
of finite, rather than infinite size. Consequently, we must assign to 
$\Gamma$ a finite value. Thus, let us assume  that $\Gamma > 0$ and 
that $\Gamma \ll \gamma \ll 1$. Note that $\gamma \ll 1$ is generated 
by $r \ll 1$. The condition $\Gamma < \gamma$ is made necessary by the 
request that the truncation of the fat tail of the subordinating 
function leaves some sign of the system complexity. However, this 
condition
implies a departure from the pure ML relaxation function, and the
form of $\Psi_{SP}$ on the time scale at which we observe the histogram 
and SP. In the short-time regime, $\gamma \ll u \ll 1$, it is 
impossible to neglect the
first term in the denominator of Eq.~\eqref{survival_ML}, and therefore,
considering that $\Gamma \ll u$
\begin{equation}
   \hat{\Psi}_{SP}(u) = \frac{1}{u + \gamma^\alpha u^{1- \alpha}},
\end{equation}
which is the Laplace transform of a stretched exponential.
On the contrary, if the condition $\Gamma \ll u \ll \gamma$ applies, the first
term in the denominator of Eq.~\eqref{survival_ML} can be neglected, and the
expression for the SP reads
\begin{equation}
   \hat{\Psi}_{SP}(u) = \frac{1}{\gamma^\alpha u^{1 - \alpha}},
\end{equation}
which, thanks to the Tauberian theorem, is the Laplace transform of an inverse
power law SP. In this condition, the fat ML function tail becomes 
visible. If $\Gamma \lesssim \gamma$, namely, $\Gamma$ is moderately 
smaller than $\gamma$, Eq.~\eqref{survival_ML} shows that the
inverse power law never appears.

Let us now find analytically the form of the subordination function
$\psi(t)$. If we adopt the expression for the memory kernel of
Eq.~\eqref{memory_kernel}, we obtain
\begin{equation}
\label{modified_laplace}
   \hat{\psi}(u) = \frac{\hat{\Phi}(u)}{u + \hat{\Phi}(u)} = \frac{1}{1 +
   \frac{u}{\hat{\Phi}(u)}} = \frac{1}{1 +  \frac{u}{\chi (u + 
\Gamma)^{2 - \mu}}}.
\end{equation}
Again, two regimes clearly appear depending on the parameters involved in
Eq.~\eqref{modified_laplace}: if $\Gamma \ll u \ll 1$, then
Eq.~\eqref{modified_laplace} becomes
\begin{equation}
   \hat{\psi}(u) \simeq 1 - \frac{1}{\chi}u^{\mu -1}
\end{equation}
that is the Laplace transform of an inverse power law distribution density. If
we explore the regime $u < \Gamma$, Eq.~\eqref{modified_laplace} reads
\begin{equation}
   \hat{\psi}(u) \simeq \frac{1}{1 + \frac{u}{\chi \Gamma^{2-\mu}}}
\end{equation}
that is the Laplace transform of an exponential distribution density, implying
that the subordination function $\psi(t)$ is an inverse power law, truncated
exponentially in the long time limit. 
  
  In conclusion, in the 
case of systems of finite size the fat tail of the subordination 
function is truncated. In the case where $r$ is not very small, and 
consequently $\gamma$ is only moderately small, the stretched 
exponential regime is not very extended and the adoption of a 
truncated subordination function generates a $\Psi_{SP}(t)$, with a 
distinct inverse power law tail. If, $r \ll 1$, and consequently the 
stretched exponential regime is very extended, a truncated 
subordination function may have the effect of canceling the inverse 
power law tail of the SP, and the stretched exponential remains the 
only sign of complexity. However, the subordination function has an 
inverse power law nature, and its power index $\mu$ is derived from 
$\alpha$ through Eq.~\eqref{relation_mu_alpha}, thereby yielding 
Eq.~\eqref{conjecture2}.

\begin{acknowledgments}
We warmly thank Dr. Marco Buiatti for interesting discussions, and for
communicating to us his results on the analysis of EEG data prior to the
publication. SB, MI and PG thankfully acknowledge Welch and ARO for financial
support through Grant no. B-1577 and no. W911NF-05-1-0205, respectively.
We are also very grateful to two unknown referees for their 
invaluable help.  We are grateful  to the first for forcing us to 
examine with critical mind our results, with the result of 
reinforcing our conviction that the properties found are a 
consequence of the cooperative processes behind the cerebral 
activities. We warmly thank the second referee for his/her 
encouragement to spell out the complexity matching between music and 
human brain. The data were analyzed as part of an institutionally approved
study of archived EEG data. 
\end{acknowledgments}

\end{document}